\documentclass[conference]{IEEEtrans}

\usepackage{sty/common}
\usepackage{sty/defs}

\IEEEoverridecommandlockouts
\IEEEaftertitletext{\vspace{-10pt}}

\title{\textbf{\LARGE Online Payments by Merely Broadcasting Messages}\\\Large(Extended Version)\thanks{Author names appear in alphabetical order, grouped by affiliation. This is an extended version of a conference article, appearing in the proceedings of the 50th IEEE/IFIP Int. Conference on Dependable Systems and Networks (DSN 2020). This work has been supported in part by the European grant 862082, AT2 -- ERC-2019-PoC, and in part by a grant from Interchain Foundation.}}

\author{
Daniel Collins, Rachid Guerraoui, Jovan Komatovic,\authorcr Matteo Monti, and Athanasios Xygkis\\EPFL \and Matej Pavlovic\\IBM Research
    \and Petr Kuznetsov\\LTCI, T\'el\'ecom Paris\\Institut Polytechnique Paris
    \and Yvonne-Anne Pignolet\\DFINITY
    \and Dragos-Adrian Seredinschi\\Informal Systems
    \and Andrei Tonkikh\\National Research University\\Higher School of Economics
}

\date{}

\begin{document}
\maketitle
\thispagestyle{plain}
\pagestyle{plain}


\begin{abstract}


We address the problem of online \emph{payments}, where users can transfer funds among themselves.
We introduce \emph{\sysname}, a system solving this problem efficiently in a decentralized, deterministic, and completely asynchronous manner.
\sysname builds on the insight that consensus is unnecessary to prevent double-spending.
Instead of consensus, \sysname relies on a weaker primitive---Byzantine reliable broadcast---enabling a simpler and more efficient implementation than consensus-based payment systems. 

In terms of efficiency, \sysname executes a payment by merely broadcasting a message.
The distinguishing feature of \sysname is that it can maintain performance robustly, i.e., remain unaffected by a fraction of replicas being compromised or slowed down by an adversary.
Our experiments on a public cloud network show that \sysname can achieve near-linear scalability in a sharded setup, going from $10K$ payments/sec (2 shards) to $20K$ payments/sec (4 shards).
In a nutshell, \sysname can match VISA-level average payment throughput, and achieves a $5\times$ improvement over a state-of-the-art consensus-based solution, while exhibiting sub-second $95^{th}$ percentile latency.

\end{abstract}


\vspace{-5pt}
\section{Introduction}
\label{sec:intro}


Online payment systems promise secure financial transactions despite distrustful parties. Transactions need to be processed correctly despite crashes and even Byzantine (i.e, malicious) behavior of a fraction of the participants~\cite{la82byzantine}.
Popular examples of payment systems include centralized solutions such as PayPal or VISA, as well as decentralized ones like Bitcoin~\cite{nakamotobitcoin} and Ethereum~\cite{ethereum}.
Numerous newer alternatives are also appearing, claiming new grounds in terms of performance or security~\cite{abr16solida,hyperledger,gilad2017algorand,gol18sbft}.


While many payment systems~\cite{hyperledger,ethereum} allow for more general transactions (known as smart contracts)~\cite{DBLPClackBB16},
in this paper we focus exclusively on \emph{payments}: allowing a participant Alice to transfer funds to a beneficiary Bob if Alice's balance is high enough. Payments represent the largest application of blockchains today, they have driven blockchain systems from their very beginning (Bitcoin) and continue to do so (Facebook's Libra and many others~\cite{danezis2015centrally,expert19forbes,guer19cryptonumber,gup16nonconsensus,kasp19,mccorry16towards,po16bitcoin,ERC20}).

We introduce \emph{\sysname}, a decentralized payment system capable of matching the performance of the largest centralized solutions (e.g., $65K$ peak, $7K$ average transactions per second, as recently reported by VISA~\cite{visa}) for payments. 


\sysname provides honest participants with \emph{robust} performance, namely stable throughput and latency; this holds independently of network scheduling (i.e., asynchrony) and of compromised replicas, as long as no more than 1/3 of the replicas are affected.
Systems building on total order (i.e., agreement), in contrast, are often susceptible to throughput degradation due to a single slow replica, typically the leader.
This is an issue that received significant attention in the literature~\cite{amir2010prime,bess14state,clement09making,du18beat,mil16honeybadger}, which we discuss in detail (\Cref{sec:related}) and also quantify experimentally (\Cref{sec:results-robust}).

An important insight underlying \sysname is that totally ordering all payments can be avoided.
Indeed, recent theoretical results show that total order (and hence consensus) is not necessary for preventing double-spending~\cite{guer19cryptonumber,gup16nonconsensus}.
The main contribution of this paper is to apply this insight by building, for the first time, an asynchronous deterministic payment system that is decentralized and consensus-free, and reporting on the empirical evaluation of this system. 


Roughly speaking, instead of requiring a total order, \emph{we give clients direct control over (the ordering of) the payments they initiate.}
Prior solutions require agreement---usually via an expensive consensus protocol~\cite{antoni18smr,FLP85,vuko15quest}---on the order across the payments of \emph{all} clients. 
Each client in \sysname independently orders their payments, thus maximizing the degree of concurrency and improving efficiency.
As a result, a payment operation essentially reduces to broadcasting a message.
A weak broadcast primitive, called Byzantine reliable broadcast (\brbroadcast) is sufficient for this purpose~\cite{bra87asynchronous,guer18at2,gup16nonconsensus}.
This primitive can be implemented in an asynchronous network, unlike consensus and total order broadcast~\cite{FLP85}.
The performance of \sysname, even in uncivil executions, is only limited by the speed of honest participants.

To record payment operations, \sysname maintains a log separately for each client.
Whenever Alice makes a new payment, she announces---through the broadcast layer---her intent to record this payment in her (replicated) log.
Payments in her log are ordered by sequence numbers she assigns herself.
\sysname guarantees that only Alice, the \emph{spender}, may record new payments in her log; we call this abstraction an \lognamelong, or \logname for short.

Essentially, preventing Alice from double-spending means preventing her from reusing sequence numbers.
To do so, the broadcast layer in \sysname provides Byzantine resilience.
This ensures that a malicious client cannot broadcast two different payments with the same sequence number.
For example, Alice cannot broadcast a payment \emph{a} for beneficiary Bob with sequence number \emph{s}, and for that same sequence number, announce a different payment \emph{a'} for beneficiary Carol.
At most one of these conflicting payments passes through the broadcast layer.
As a result, Alice cannot double-spend.

\sysname\ distinguishes between \emph{clients} of the system and \emph{replicas} that operate the payment system.
Clients usually connect to the system infrequently to submit payments and  check their balance.
Intuitively, each client is a lightweight participant and thus relies on a certain replica---called a \emph{representative}---to broker her payments via broadcast. Nevertheless, each client controls the ordering of her own payments.
Replicas maintain the system state (i.e., client \lognameplural), remain well-connected to each other, and implement the broadcast-based replication layer.
Payments are safe and live as long as the spender and 2/3 of the replicas, including the representative replica handling the request, are correct. 

This distinction between \emph{client} and \emph{replica} allows the number of clients in \sysname to scale independently of replicas; a client may, of course, be its own representative.
The broadcast layer (implemented by replicas) relies on quorum systems~\cite{malk97byzantinequorums} to ensure Byzantine resilience, and consequently does not scale beyond tens or hundreds of replicas.
The number of clients, on the other hand, can be orders of magnitude larger.


For pedagogical reasons, we proceed in an incremental manner. We first discuss an implementation of \sysname{} without using digital signatures, before moving to a more efficient scheme with digital signatures and fewer messages. 
To scale the number of replicas in \sysname, we employ a sharding scheme: We partition the system state and replicate each partition among a subset of replicas.
Sharding a payment system is difficult if payments need to be totally ordered (i.e., based on consensus):
Approving a cross-shard payment requires all involved shards to coordinate, usually via a  2PC protocol~\cite{kok18omniledger,zama18rapidchain}.
We sidestep this major difficulty because \emph{the shard of the spender can---in our case---unilaterally approve a cross-shard payment}.
\sysname requires no cross-shard coordination on the critical path of payment execution.
The beneficiary receives her funds via an asynchronous notification mechanism after the spender's shard approves it.
Again, for simplicity of presentation, we present first the non-sharded case before explaining the sharded solution.

We evaluate \sysname on a public wide-area cloud network (Amazon EC2). 
We show that even without sharding and even in synchronous and failure-free executions,  \sysname outperforms a state-of-the-art consensus-based payment system.
Considering four shards with $52$ replicas per shard, \sysname can sustain up to $20K$ payments per second at sub-second (95$^{th}$ percentile) latency.
But more importantly,
\sysname  provides \emph{robust performance}: In executions where some replica crashes or suffers from high network latencies, overall throughput  is unaffected (except for the failed replica).
Leader-based consensus systems can experience throughput degradation in such situations, to the point where payment execution blocks altogether when the leader is affected, as we show empirically.


\vspace{.15cm}\noindent\textbf{Contributions.}
We design \sysname with a focus on \emph{payments} for a \emph{permissioned} model.
Our system lacks some capabilities compared to mature blockchains (e.g., Sybil resistance, smart contracts, or full decentralization as Bitcoin or  Ethereum) or global payment systems (e.g., negative balance, fraud detection as VISA).
We do not intend \sysname to replace such systems, but rather demonstrate the efficiency and power of broadcast for improving existing solutions.

\sysname circumvents consensus-inherent complexities, being the first payment system that is completely asynchronous, deterministic, and guarantees robust performance.
In summary:
\begin{compactenum}
    \item \sysname introduces the abstraction of an \textbf{\lognamelong}: A record of client payments uniquely controlled by a certain client. \sysname maintains the consistency of \lognamelongplural through a \textbf{weak broadcast primitive}, thus maximizing concurrency and efficiency.
    
    \item \sysname is fully \textbf{asynchronous}, including support for an asynchronous sharding mechanism for scalability.
    
    
    \item Our \sysname implementation can match the performance, with respect to payments, of centralized solutions (e.g., VISA) in a \textbf{robust} manner. 
\end{compactenum}


The rest of this paper is organized as follows.
We first overview \sysname (\Cref{sec:motivation}) and then detail its payment protocol (\Cref{sec:payment-protocol}).
We describe our two implementations of \sysname (\Cref{sec:system-versions}), and present our asynchronous sharding (\Cref{sec:sharding-mechanism}) scheme.
Then we discuss a thorough experimental evaluation of \sysname (\Cref{sec:eval}) and present related work (\Cref{sec:related}).
In the appendix of this paper, we provide additional details on asynchronous reconfiguration (\Cref{sec:reconf-appendix}) and the broadcast layers of \sysname (\Cref{appendix:brb-echo}).


\vspace{-2pt}
\section{Overview}
\label{sec:motivation}


At the heart of \sysname lie two building blocks that are closely related to each other. These distinguish our payment system from prior solutions, namely: (1) \lognamelongplural, or \lognameplural, and (2) a broadcast-based replication layer.

\vspace{.15cm}\noindent\textbf{Exclusive Logs.}
An \logname is an append-only log comprising all the \emph{outgoing} payment operations initiated by a certain client.
Intuitively, the \logname of Alice can be seen as her personal ledger of expenditures.
Alice is exclusively allowed to append payments to her \logname, and we refer to Alice as the \emph{owner} of her log.

It is Alice herself who establishes the ordering of payment operations in her \logname, by assigning a sequence number to each payment.
Besides a sequence number, each payment also specifies the \emph{spender} (which is always Alice in this case), the \emph{amount}, and the \emph{beneficiary} of the payment.

\begin{figure}[t]
\centering
\vspace{-.5cm}
\includegraphics[width=0.35\textwidth]{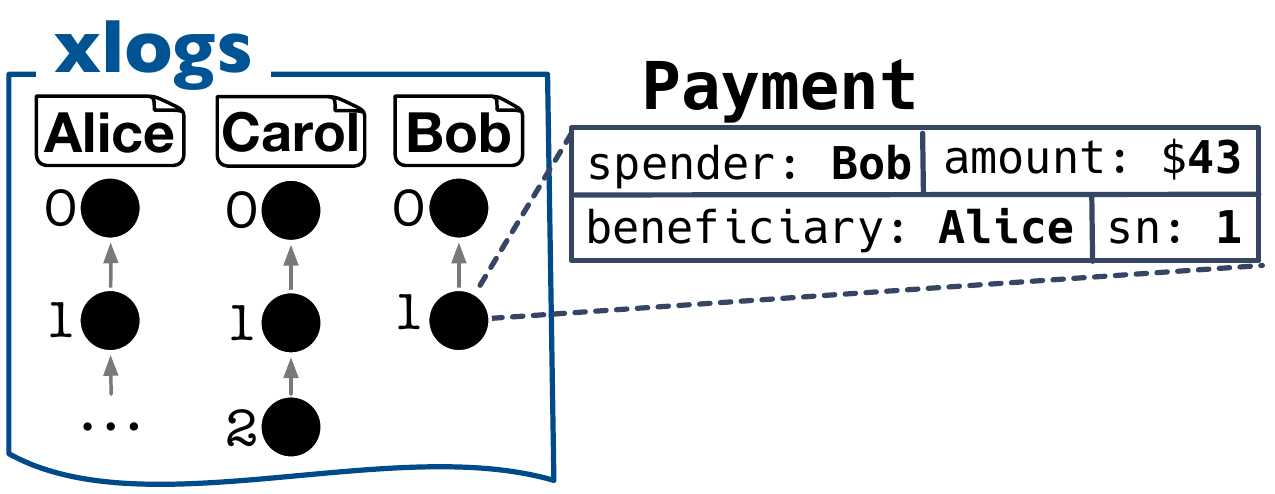}
\vspace{-.1cm}
\caption{\textbf{System state in \sysname, consisting of multiple \lognameplural (\lognamelongplural)}.
Each \logname contains payments operations having the same spender (i.e., belonging to the same client). For example, Bob's \logname comprises two operations; the second of these is a payment of $\$43$ from Bob to Alice, having sequence number $1$.}
\label{fig:histories}
\vspace{-.4cm}
\end{figure}

\sysname's state consists of multiple \lognameplural, one per client, as we sketch in~\Cref{fig:histories}.
In the basic version of \sysname, each replica holds a copy of the entire state (we revise this to consider sharding in \Cref{sec:sharding-mechanism}).

In a static system, storing \lognameplural\ could be completely avoided, by only storing balances and a single sequence number for each client.
Storing the \lognameplural is crucial for reconfiguration of \sysname, i.e., for dynamically changing system membership (\Cref{sec:reconf-appendix}) and to enable auditability.





\vspace{.15cm}\noindent\textbf{Consistent Replication of \lognameplural.}
The goal of the replication layer in \sysname is to keep all \lognameplural consistent across replicas despite Byzantine failures.
To do so efficiently, we exploit an idiosyncrasy of \lognameplural, namely that each such log restricts append access to the (authenticated) owner client.
Consequently, we never have to deal with concurrent modifications on a \logname. Each client can modify their own \logname autonomously:  \sysname supports concurrent modification of any number of \lognameplural.

Each client is associated with a single replica acting as its \emph{representative}.
A single replica can represent many clients.
The representative is in charge of broadcasting the client's payments to other replicas, and corresponds to a \emph{broker} or a \emph{bank}.
Akin to a real bank, only the representative can broadcast outgoing payments for a client's \logname .
All payments still have to be ordered and submitted by the client.
Unlike with banks, however, multiple replicas in \sysname replicate each client's data (\logname).

\begin{figure}[t]
\centering
\vspace{-.5cm}
\includegraphics[width=.70\columnwidth]{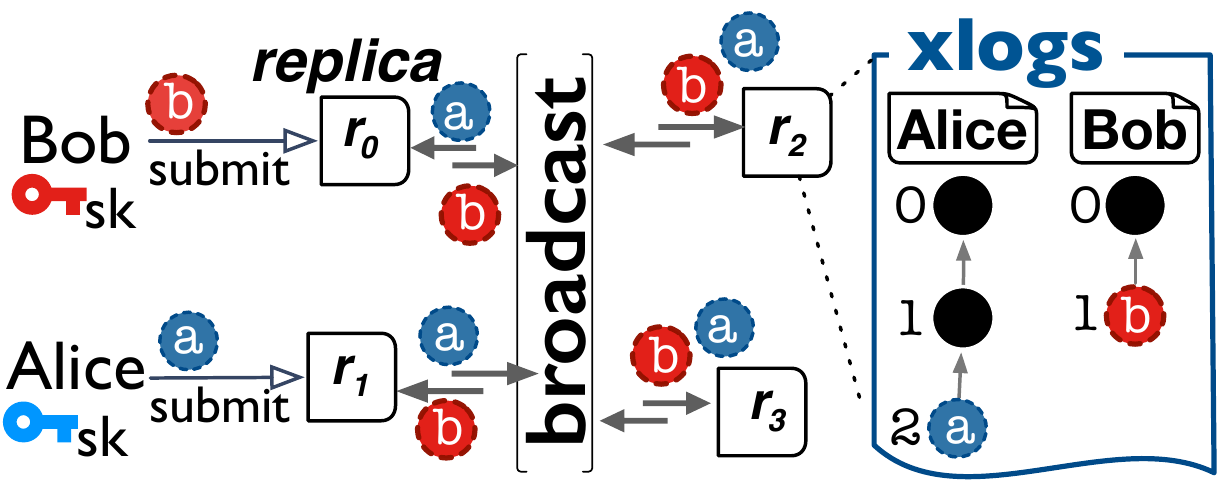}
\vspace{-.1cm}
\caption{\textbf{Payment protocol overview in \sysname.} When Alice wants to make a payment \emph{a}, she simply submits \emph{a} to her representative replica $r_1$. This replica handles the broadcasting of Alice's payment. Eventually, all correct replicas deliver \emph{a}, append this payment to Alice's \logname (on position 2), and update client balances accordingly to reflect this payment.}
\label{fig:high-level}
\vspace{-.4cm}
\end{figure}

A client performs a payment by submitting it to  her representative \emph{r}.
Replica \emph{r} ensures that all copies of the client's \logname are updated consistently.
To this end, replicas implement a broadcast primitive guaranteeing the following crucial property: no client can announce two conflicting payments (i.e., with the same spender) for the same sequence number, despite Byzantine clients and/or replicas.
In other words, \sysname guarantees total order within---but not across---\lognameplural, departing from prior designs that employ a total order across all payments (\Cref{fig:high-level}). From the clients' perspective, \sysname provides FIFO guarantees~\cite{hu10zookeeper,llo11cops}.

As we pointed out, current decentralized payment systems achieve consistent replication by executing a consensus protocol~\cite{hyperledger,kok18omniledger,nakamotobitcoin},
while also tackling broader problems (e.g., implementing smart contracts).
In many cases, consensus poses a performance bottleneck and is the usual suspect in problems regarding correctness or complexity~\cite{abrah17revisiting,cac17blwild,clement09making}, given its numerous impossibilities and inherent tradeoffs~\cite{antoni18smr,FLP85,guerraoui2019100,mil16honeybadger,vuko15quest}.

In \sysname, we replace the consensus building block with a broadcast layer.
Formally, \sysname builds on Byzantine reliable broadcast (\brbroadcast). 
This should not be confused with classic Byzantine Agreement (\textsc{ba}), which is \emph{unsolvable} in the asynchronous model we assume~\cite{FLP85}.
The \brbroadcast primitive is not novel, appearing in the literature for over $30$ years, starting with Bracha \& Toueg~\cite{bra87asynchronous,br85acb}.
The crucial difference to \textsc{ba} that allows asynchronous implementations of \brbroadcast\ is termination: \textsc{ba} always guarantees termination, whereas \brbroadcast\ does not guarantee this property if the spender is faulty~\cite{guer19cryptonumber}.
Stated differently, if the spender client proposes two conflicting payments (double-spending) under \brbroadcast, it is possible that no payment will ever execute. 







\vspace{-2pt}
\section{Payments in \sysname}
\label{sec:payment-protocol}



\sysname is a replicated system running on $N$ \emph{replicas} of an asynchronous network.
The replicas implement a broadcast-based replication layer and maintain the full system state, which they update consistently to reflect client payments.
Both clients and replicas hold an identifying public/secret key-pair. 
We assume that \emph{(1)} replica key-pairs are distributed in advance among all replicas, which makes \sysname a permissioned payment system, and  \emph{(2)}  the mapping of clients to their representative replicas is publicly known.
We assume less than $N/3$ of replicas to be Byzantine.
This is a standard assumption, but we revisit this aspect later, when we introduce partial replication via sharding~(\Cref{sec:sharding-mechanism}).
We now describe the basic payment protocol.

\noindent At a high level, payment execution comprises three steps:
\begin{compactenum}
\item The client \textbf{submits} a payment \emph{a} to her representative.
\item The representative \textbf{broadcasts} \emph{a} to all replicas.
\item Replicas locally \textbf{approve} payment \emph{a} and append it to their local copy of the corresponding client's \logname.
\end{compactenum}
If the client and representative replica are both correct, each of these three steps is guaranteed to terminate.
A correct client, however, is unaffected by other Byzantine clients.
Specifically, no client will ever be able to double-spend or prevent any other client from performing payments, as long as less than $1/3$ of replicas are malicious. 

We now describe the three aforementioned steps in detail.
For presentation simplicity, we use pseudocode inspired by Golang which we assume to execute atomically.

\vspace{.15cm}\noindent\textbf{Submitting a Payment.}
In~\Cref{lst:start-payment} we describe the algorithm a client Alice implements to submit a payment.
First, she creates a payment message comprising the identity of the spender (herself), the sequence number she assigns to this payment, as well as the identity of the beneficiary,
and the amount.
Alice then increments her sequence number, and finally sends the payment to her representative replica through an authenticated channel (line~\ref{line:broadcast-request}).

\begin{lstlisting}[
  caption = {\textbf{Client Alice submits a new payment.}},
  label={lst:start-payment},
  language=Astro,
]
@executes at spender Alice
@local state: Client Alice;
              Sequence number mySN;
func Pay(Client b, Amount x):
  a := <Alice, mySN, b, x>(*@\label{line:payment-details}@*)
  mySN += 1 // Increment our sequence number. (*@\label{line:seqnr}@*)
  Send(a) // Submit the payment to her representative. (*@\label{line:broadcast-request}@*)
\end{lstlisting}
\vspace{-5pt}

\vspace{.15cm}\noindent\textbf{Broadcasting a Payment.}
When the representative receives Alice's payment, it broadcasts this payment among the replicas using the underlying Byzantine reliable broadcast (\brbroadcast) layer. \brbroadcast ensures that all correct replicas will eventually deliver Alice's message if her representative replica is correct.
This layer implements
a \emph{consistency check} ensuring that no two correct replicas deliver a different message for the same sequence number of a certain client.
We discuss the broadcast layer in more detail later (\Cref{sec:system-versions}).

\vspace{.15cm}\noindent\textbf{Approving a Payment.}
Upon delivery of a payment message from the broadcast layer, each replica locally \emph{approves} Alice's payment, and then \emph{settles} it (see lines~\ref{line:approve},~\ref{line:settle} in~\Cref{lst:broadcast-deliver}).

\begin{lstlisting}[
  caption = {\textbf{A payment \emph{a} is ready}. Each correct replica runs this callback upon delivering \emph{a} from the underlying broadcast layer.},
  label={lst:broadcast-deliver},
  %float,
  firstnumber=last,
  language=Astro,
]
@executes at all system replicas
@local state: SeqNrMap    sn[..] // last SN per client
              BalancesMap bal[..] // balances per client (*@\label{line:bal}@*)
              XLogMap     xlogs[..] // xlogs of clients
callback Deliver(a) (*@\label{line:broadcast-deliver}@*)
  approve(a) // Blocks waiting for approval of this payment(*@\label{line:approve}@*)
  settle(a)(*@\label{line:settle}@*) // Apply the payment locally
\end{lstlisting}

\noindent\textit{Approval.} The approval procedure is described in \Cref{lst:approve-tot}.
Each replica in \sysname executes this procedure with the goal of ensuring two important properties:
\begin{compactenum}
\item All Alice's preceding payments are approved (line~\ref{line:wait-sn}).
\item Alice has sufficient funds for her payment, as reflected by her balance (line~\ref{line:wait-balance}).
\end{compactenum}

If both Alice and her representative are correct, these conditions may be unfulfilled at replica \emph{q} only if \emph{q} has not yet approved either:
\begin{compactenum}
\item Alice's preceding payment, or
\item Some other payment crediting Alice.
\end{compactenum}
In such a case, \emph{q} simply waits until both conditions are satisfied.
Under normal conditions, correct clients would initiate payments which they can fulfill straight away.
Nevertheless, it can be useful to allow Alice to initiate payments despite not having enough funds to settle them right away.
Such payments (and all subsequent ones) will not be approved until Alice has sufficient balance.

\begin{lstlisting}[
  caption = {\textbf{Payment approval.} Every replica executes this to approve a payment \emph{a}, assuming spender Alice.},
  label={lst:approve-tot},
  firstnumber=last,
  language=Astro,
]
func approve(a)
  let a be <Alice, n, _, x>
  wait until sn[Alice] = n - 1 // Approval criterion (1) (*@\label{line:wait-sn}@*)
  wait until bal[Alice] (*@$\ge$@*) x // Approval criterion (2) (*@\label{line:wait-balance}@*)
\end{lstlisting}

\noindent\textit{Settling.} As the final step in payment execution, each replica \emph{settles} this payment (\Cref{lst:settle-tot}), i.e., updates the balances of the spender and beneficiary, updates the sequence number of the spender client, and records the payment in the spender's \logname.
Note that maintaining the whole history of payments in the \logname is not strictly necessary for the safety of the basic payment protocol. 
In a \emph{static} system, storing the balances and sequence numbers for each client suffices. Yet, having this log enables auditability and supports a system where the set of replicas may change for growth, repair or reconfiguration~(\Cref{sec:reconf-appendix}).





\begin{lstlisting}[
  caption = {\textbf{Payment settling procedure.} Each replica executes this protocol to transition a payment \emph{a} to the final, settled state.},
  label={lst:settle-tot},
  firstnumber=last,
  language=Astro,
]
func settle(a)
  let a be <Alice, n, b, x>
  bal[Alice] -= x // Withdraw from Alice's balance (*@\label{line:withdraw}@*)
  bal[b] += x // Deposit to beneficiary (*@\label{line:deposit}@*)
  sn[Alice] += 1
  xlogs[Alice].append(a)
\end{lstlisting}

\vspace{-.15cm}\noindent\textbf{Client notification.}
By default, we assume clients to be lightweight and intermittently connected, so we omit a specific step of notifying clients that their transaction settled (or is cleared in the system).
It is simple, however, to achieve end-to-end notification, by having the client query her representative for the status of the payment. The latter can reply after it has finished with the \emph{settle} step.

\vspace{.1cm}\noindent\textbf{Checking the Balance.}
A client can check her balance by querying her representative \emph{r}. 
To obtain the balance, replica \emph{r} simply returns the value from the \emph{bal} state (defined on line~\ref{line:bal},~\Cref{lst:broadcast-deliver}). 

\section{A Tale of Two Versions}
\label{sec:system-versions}

We now turn our attention to the broadcast layer in \sysname.
Replicas use this layer to replicate client payments consistently, and it is implemented using a \brbroadcast protocol.
The \brbroadcast interface has two methods.
First, a replica \emph{r} can use \emph{Broadcast(a)} to reliably send payment \emph{a} to all replicas in the system.
Second, the \emph{Deliver(a)} callback triggers at any correct replica to notify about the delivery of payment \emph{a}.
The broadcast layer is aware of the payload \emph{a}, which specifies: the spender \emph{s}; sequence number \emph{n}; beneficiary \emph{b}; and amount \emph{x}.
We denote the pair \emph{(s,n)} to be the \emph{identifier} of payment \emph{a}.
We now define the properties of the broadcast layer, inspired by \cite{ma97secure}, where payment identifiers are particularly important:
\begin{compactitem}
\item \emph{Agreement}. If a correct replica delivers a payment \emph{a} with identifier \emph{(s, n)}, then no correct replica delivers a payment \emph{a'} $\ne$ \emph{a} with the same identifier.
\item \emph{Integrity}. A correct replica delivers a payment \emph{a} at most once, and under the condition that \emph{a} is broadcast by a replica \emph{r}.
\item \emph{Reliability}. If the broadcaster replica of payment \emph{a} is correct, then all correct replicas eventually deliver \emph{a}.
\item \emph{Totality} (optional). If a correct replica delivers payment \emph{a}, then every correct replica eventually delivers \emph{a}.
\end{compactitem}

There is a rich history of protocols implementing \brbroadcast~\cite{bra87asynchronous,cac11intro,cach02sintra,MR97srm}.
We mark totality property as optional because there exist \brbroadcast protocols which in fact do not offer this property by default.
Such protocols are appealing because they are more efficient.
If totality is missing, however, an adversary can mount a \emph{partial payments} attack against our payment protocol, as follows.
Suppose Alice issues a payment to Bob, who initially has \$0.
Let Alice's representative $r_{A}$ be malicious, whereas the representative $r_{B}$ of Bob is correct.
In the absence of totality, since $r_{A}$ is malicious, only $r_{B}$ would deliver and settle Alice's payment, while Bob's \logname in any other replica still has a balance of \$0. 
Bob cannot spend the \$10 he received, because there are no $2f+1$ replicas with the updated version of Bob's \logname. 

We implement and evaluate two versions of \brbroadcast , and thus obtain two versions of our system: \sysnameecho and \sysnamesig.
\sysnameecho uses a \brbroadcast protocol \cite{bra87asynchronous} that has a similar communication pattern to our consensus-based baseline and allows for a fair performance robustness comparison (\cref{sec:results-robust}).
\sysnamesig, on the other hand, uses stronger cryptographic primitives to reduce communication complexity, achieve higher performance, and enable sharding.
Additionally, \sysnamesig lacks the totality property, so we compensate for that with an additional mechanism to prevent the attack we mentioned above.

Both \brbroadcast protocols underlying \sysnameecho and \sysnamesig assume less than a third of replicas to be Byzantine and offer the API we specified earlier.
We now describe the broadcast protocols in our systems; for the pseudocode, we refer the interested reader to the appendix (\Cref{appendix:brb-echo}).

\subsection{Broadcast Protocols \& \sysname Versions}
\label{sec:variants-details}



\vspace{.15cm}\noindent\textbf{\sysnameecho} implements \brbroadcast based on Bracha's algorithm~\cite{br85acb}.
Let \emph{a} be a payment with identifier \emph{(s, n)} that the representative replica \emph{r} is broadcasting on behalf of spender client \emph{s}.
This protocol relies on authenticated links, e.g., via message authentication codes (MACs), and comprises three phases.

\noindent (1) \opname{Prepare}. To broadcast payment \emph{a}, correct replica \emph{r} simply sends \emph{a} to all replicas in the system.

\noindent (2) \opname{Echo}. The first time a replica \emph{q} receives a payment with identifier \emph{(s, n)}, it sends an \opname{Echo} message for this payment to all replicas in the system.

\noindent (3) \opname{Ready}. In this last phase of the protocol, every replica \emph{q} waits to collect a Byzantine quorum~\cite{malk97byzantinequorums} of \opname{Echo} messages for tuple \emph{(s,n)} and then \emph{q} sends a \opname{Ready} message.
Alternatively, replica \emph{q} may send a \opname{Ready}  after observing $f+1$ \opname{Ready} messages.
A correct replica delivers payment \emph{a} after gathering $2f+1$ matching \opname{Ready} messages for \emph{a} and after having delivered the previous payment of client \emph{s}, i.e., the payment with identifier \emph{(s, n-1)}.





Observe that Bracha's protocol entails two phases (\opname{Echo} and \opname{Ready}) of all-to-all communication, i.e., has message complexity of $O(N^2)$.
On the plus side, this protocol uses MACs, thus it is not computationally intensive.

\vspace{.15cm}\noindent\textbf{\sysnamesig} implements the broadcast layer using a \brbroadcast protocol with linear ($O(N)$) message complexity.
At a high-level, this protocol employs digital signatures, and also comprises three phases.
The first phase, called \opname{Prepare}, is identical to the first phase of the broadcast protocol of \sysnameecho.
The other two phases of this protocol are as follows:

\noindent (2) \opname{Ack}.
Upon receiving payment \emph{a} from replica \emph{r}, every replica \emph{q} verifies whether there exists \emph{a'} $\ne$ \emph{a} previously received for identifier \emph{(s, n)}.
If this is not the case, then \emph{q} sends a signed \opname{Ack} message (i.e., a signed hash) of \emph{a} directly to replica \emph{r}.
Otherwise, replica \emph{q} does nothing.

\noindent (3) \opname{Commit}.
Upon gathering a Byzantine quorum~\cite{malk97byzantinequorums} of matching acknowledgments for payment \emph{a}, replica \emph{r} sends to all other replicas a \opname{Commit} message, comprising the gathered acknowledgments.
Each correct replica delivers \emph{a} after receiving a correct commit message for \emph{a}.

To prevent the partial payments attack, we introduce \textit{dependencies} in \sysnamesig.
A correct replica that approved Alice's payment, unicasts the signed approval called \opname{Credit} message to Bob's representative, and allows Bob to prove the existence of a payment crediting his account unequivocably with $f+1$ such \opname{Credit} messages. To this end, Bob's representative replica collects and aggregates \opname{Credit} messages for the same incoming payment into a dependency certificate for Bob's \logname. 
If Bob's representative fails in any way, this certificate is not lost; the certificate is permanently stored as \opname{Credit} messages, distributed across the replicas that approved the payment, so it can be reconstructed directly from these replicas.

Note that replicas must keep track of used certificates, ensuring that each payment takes effect not more than once. This way, it is impossible for replicas to mistakenly apply a dependency twice (e.g., double-deposit, as in a replay attack).
Listings \ref{lst:approve-tot} and \ref{lst:settle-tot} have to be adjusted to take dependencies into account, see pseudocode in~\Cref{appendix:brb-echo}.

Certificates also play an important role in a sharded environment, as they are transferable across shards: They enable Bob to spend the money mentioned in the dependency not only within his representative's shard, but also across shards (\Cref{sec:sharding-mechanism}).
Whenever Bob submits an outgoing payment, his representative replica attaches the accumulated dependencies alongside the outgoing payment.


\vspace{.15cm}\noindent\textbf{Comparison.}
\sysnamesig is well-suited for environments where bandwidth is scarce (e.g., WAN), whereas \sysnameecho has lower computation requirements and is therefore suited for systems where computing resources are more scarce.
Given a batching scheme, however, we can amortize the cost of digital signatures in \sysnamesig, as we describe later (\Cref{sec:eval-systems}).
Moreover, we expect the typical deployment of our system to be a wide-area network where bandwidth is the scarce resource.
Because of these reasons, \sysnamesig has an edge over \sysnameecho in terms of performance---a hypothesis we quantify in our experimental evaluation (\Cref{sec:results-perf}).

The two systems handle transitive transactions differently. 
\sysnameecho does not reject insufficiently funded transactions (line~\ref{line:wait-balance},~\Cref{lst:approve-tot}), instead it queues them until enough funds arrive. 
Queuing is necessary even with totality, since different replicas may receive crediting transactions at different times. 
Instead, the dependencies mechanism in \sysnamesig allow the spender’s representative to prove that the spender has sufficient funds to issue a payment.

There is an additional important distinction between \sysnameecho and \sysnamesig: the latter is amenable to sharding.
To understand why this is the case, we observe that sharding requires the approval of payments across different shards.
In other words, some shard \emph{s1} has to convince some other shard \emph{s2} that \emph{s1} approved a certain payment and \emph{s2} can settle it.
Digital signatures simplify this transfer of trust between shards, because the payment of a spender from \emph{s1} appears as a dependency in the \logname of the beneficiary in \emph{s2}. 
Replicas in \emph{s2} accept this dependency when they verify it is signed by $f+1$ replicas of \emph{s1}.
In the next section, we provide the full details of the sharding mechanism which we implement in \sysnamesig.





\vspace{-2pt}
\section{Asynchronous Sharding}
\label{sec:sharding-mechanism}
So far we described our payment protocol (\Cref{sec:payment-protocol}) assuming full replication.
In this model, all replicas maintain a full copy of the system state (i.e., \lognameplural) and approve and settle every payment.
The full replication architecture is simple to understand and implement, and excels at small scale.
This design poses two scalability problems. First, throughput degrades with increasing replica count (as we observe experimentally in~\Cref{sec:results-perf}). Second, each replica has to keep more state as the number of \lognameplural (i.e., clients) increases.

We now refine the architecture of our payment system with sharding, which \sysnamesig implements.
Sharding is a well-known technique~\cite{ad16slicer,al2017chainspace,bezerra2014scalable,kok18omniledger,luu2016secure,wang2019monoxide,zama18rapidchain}, allowing our system to scale-out in terms of both number of replicas and number of clients.
We define a \emph{shard} as a subset of system replicas, and to be associated with a subset of all \lognameplural.
We use the notation \shardof{$\cdot$} to denote the shard to which some replica or client ``$\cdot$'' belongs.
Importantly, sharding requires strengthening our assumption from \Cref{sec:payment-protocol}, so that the threshold $N/3$ on Byzantine replicas applies to every shard.


Intuitively, each shard in \sysnamesig executes an instance of the basic payment protocol (\Cref{sec:payment-protocol}) for its associated clients. It also incorporates an additional mechanism that not only prevents partial payment attacks, but also supports sharding seamlessly.
The broadcast step of  \sysnamesig is executed in the shard of the spender, while the \opname{Credit} messages may be sent to a representative in another shard. 

\vspace{.15cm}\noindent\textbf{} 
 Let us consider a payment of amount \emph{x} from spender \emph{A} to beneficiary \emph{B} and illustrate how \sysnamesig processes it. Let \emph{r} be the representative replica of \emph{A}.




After broadcasting and approving the payment of client \emph{A}, 
all honest replicas in shard \shardof{A} 
%
unicast a \opname{Credit} message to the beneficiary's representative in \shardof{B}, indicating the crediting of amount \emph{x} to the balance of client \emph{B}.
This message comprises all details of this payment (including the sequence number \emph{n} assigned by client \emph{A}), as well as a signature \emph{sig} indicating the approval of the payment from the perspective of that replica.
 The representative of \emph{B} interprets f+1 distinct \opname{Credit} messages as a dependency certificate, i.e., a proof that the payment has been accepted by shard \shardof{A}. This dependency certificate is stored at the representative of \emph{B} and gets added to\emph{B}'s balance when the next outgoing transaction issued by \emph{B} is settled by the replicas in shard \shardof{B}.




Traditional sharded designs employ a 2PC protocol for coordinating transactions that span multiple shards~\cite{ba11megastore,glenden11scatter}.
The 2PC protocol relies on synchrony and has a delay of $3$ communication steps; each such step usually has complexity $O(m)$ and in the Byzantine case it can reach up to $O(m^2)$, where $m$ is the size of a shard~\cite{glenden11scatter,guer16atum}.
In contrast, our protocol based on the \opname{Credit} message entails exactly $1$ communication step and has overall complexity $O(m)$.
In our experiments with \sysnamesig implementing the Smallbank application~\cite{dinh17blockbench} we observe that this sharding mechanism has negligible overhead (\Cref{sec:smallbank-eval}).

The insight enabling such a simple sharding mechanism in \sysname is that we \emph{decouple payment processing at the spender from the beneficiary}.
In fact, this mechanism is orthogonal to how a payment is executed inside a shard (e.g., using a consensus or a broadcast based protocol).


\vspace{-5pt}
\section{Experimental Evaluation}
\label{sec:eval}


We now report on the experimental evaluation of our consensus-free approach to payment systems.
We first describe the systems we evaluate, namely \sysname I and II and a baseline based on consensus (\Cref{sec:eval-systems}).
We also detail our evaluation methodology (\Cref{sec:methodology}) and present  our comprehensive evaluation, covering both the common-case and performance robustness (\Cref{sec:results-perf,sec:results-robust}).

\vspace{-1pt}
\subsection{Systems under Evaluation}
\label{sec:eval-systems}

We build our baseline on top BFT-SMaRt, a mature state-of-the-art BFT SMR (i.e., consensus) implementation~\cite{bess14state}, used, for example, as the ordering service of Hyperledger Fabric~\cite{sou18byzantine}%
\footnote{
In general, there is a notable difference in complexity between consensus---in particular, the Byzantine-fault tolerant versions---and broadcast algorithms.
Both \sysname implementations require less than $3.5K$ \textsc{loc} in Golang. Contrast this with \texttt{libpaxos}~\cite{libpaxos}, a simple consensus implementation for the crash-only model, stretching over more than $6K$ \textsc{loc} in C.
At the time of its original publication, the BFT-SMaRt implementation counted around $13.5K$ \textsc{loc} in Java~\cite[\S III]{bess14state}.
}.
For both \sysname systems and BFT-SMaRt under evaluation we assume the optimal threshold of $N=3f+1$ replicas, where $f$ bounds the number of faulty replicas.


\vspace{.1cm}\noindent\textbf{Batching in \sysname I and II.}
%
%
We employ a 1- or 2-level batching scheme, depending on the variant of our system.
First, we perform batching at the level of the broadcast protocol.
Note that the first step of the broadcast protocol (\opname{Prepare} in~\Cref{sec:variants-details}) is identical across our two systems.
Briefly, some replica \emph{i} sending a \opname{Prepare} is the one assembling a batch of payments---potentially from different clients---with the goal of amortizing both the cost of message authentication and network processing overheads.

Second, to reduce the overhead of digital signatures necessary for the \brbroadcast and the \opname{Credit} messages, \sysnamesig groups together payments for which the beneficiary clients have the same representative replica.
Thus, when a replica \emph{i} builds a batch of payments to be broadcast, it includes sub-batches of payments segregated according to the beneficiary replica. As a result, there are as many signatures for \opname{Credit} messages as there are sub-batches.
All payments in the batch are processed together during broadcast, while the payments in the sub-batches are processed together when settling and unicasting.

Even though batching alleviates the computational burden of cryptographic signatures, it relies on the fact that clients have to trust their replicas for not issuing transactions without the former's consent. However, our approach can protect clients from malicious representative behavior if the same protocol adopts end-to-end client signatures.

\vspace{.1cm}\noindent\textbf{Cryptography in \sysnamesig.}
We used ECDSA on the NIST P-256 curve provided from the Golang standard library, which offers adequate performance.
To avoid cryptographic operations acting as a CPU bottleneck, we use one signature per batch of 256 payments in the broadcast layer.
With this batch size, \sysnamesig's performance is only limited by available bandwidth.




\vspace{-1pt}
\subsection{Evaluation Methodology}
\label{sec:methodology}

We use Amazon EC2 as our experimental platform.
Throughout all experiments, we use commodity-level virtual machines (VMs) of type \emph{t2.medium}~\cite{ec2}, equipped with 4GiB of RAM and 2 vCores.
Unless we explicitly state otherwise, we deploy each system so that every replica executes on a separate VM.
This avoids creating noise in our results, which could arise due to performance interference.

Our deployment setup comprises four Amazon EC2 regions in Europe, namely Frankfurt, Ireland, London, and Paris. On average, the bandwidth and round-trip latency across machines of these four regions is around 30 MiB/sec and $20ms$, respectively. 
We deploy the replicas of each system randomly across the corresponding regions.
This deployment reflects a scenario where participants are localized in one geographic region of the globe (Europe).
Later in our experiments, we also introduce network delays at each replica. As a result, we lessen the effect of sub-millisecond latency between replicas in the same region and obtain more realistic conditions with larger latencies (\Cref{sec:smallbank-eval}).


\begin{figure*}[t]
\vspace{-.5cm}
\centerline{\includegraphics[width=\textwidth]{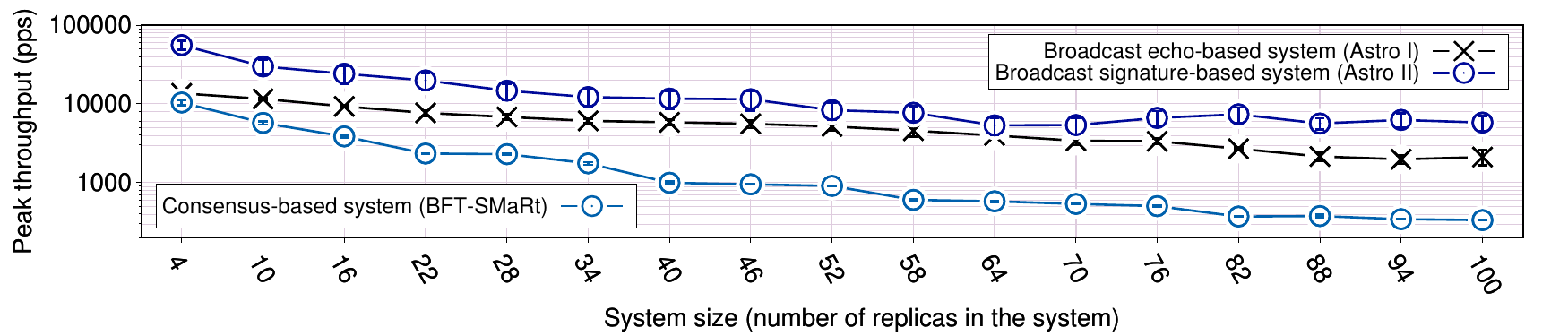}}
\vspace{-1pt}
\caption{\textbf{Throughput vs. system size.} We measure peak throughput as we increase the number of replicas in different payment system implementations, one based on consensus (BFT-SMaRt), and two based on broadcast (\sysname I and II). We do not employ sharding.}
\vspace{-.35cm}
\label{fig:peak-throughput-det}
\end{figure*}

We use up to $15$ VMs to deploy clients.
Each request from a client represents one payment.
A request contains three fields (the \emph{spender} and \emph{beneficiary} identities, along with the \emph{amount}) and the client authentication data.
The beneficiary and amount fields are random, and each payment operation covers roughly $100$ bytes.


For simplicity, we place all client VMs in Ireland.
Spreading clients around Europe does not influence our results.
Each such VM hosts a varying number of client processes.
The number of processes varies greatly, depending on each system and the system size.
For instance, to saturate BFT-SMaRt at system size $N=4$, we use around $800$ total client threads; for $N=100$, $30$ threads are sufficient for saturation.
For \sysname, we require more client threads to reach saturation, since they are capable of higher performance.
We report the maximum achievable performance: our experiments assume that all transactions can be settled immediately, i.e. clients have enough balance, so transactions can not be blocked due to insufficient funds.

For throughput, we report on how many payments each system settles per second, labeled \emph{pps}.
All experiments have a runtime of $60$ seconds, and we present the average result across $3$ runs.
We also plot the standard deviation, but often this is negligible and not clearly visible in the plots.

In BFT-SMaRt, each client keeps connections to all replicas (a design decision of this protocol)~\cite{bess14state}.
For this reason, all BFT-SMaRt clients experience similar latencies.
In our results we report on the latency as observed by a random client.
In our \sysname systems, each client connects to a single, random replica.
To make all replicas execute payments (which is the most realistic scenario), clients pick and submit their workload to a random replica.



\vspace{-1pt}
\subsection{Performance Evaluation Results}
\label{sec:results-perf}

We seek to answer the broad question of how our asynchronous approach compares in performance, at varying system sizes, with the consensus baseline.
We discuss microbenchmarks for latency and throughput in a single shard (\Cref{sec:microbenchmarks}), as well as results with the Smallbank~\cite{alomari2008cost} benchmark in a sharded scenario (\Cref{sec:smallbank-eval}).


\subsubsection{Microbenchmarks}
\label{sec:microbenchmarks}
\hfill\newline
\noindent\textbf{Throughput.}
In \Cref{fig:peak-throughput-det} we depict how throughput evolves as a function of system size.
For each system size, we plot the peak throughput, i.e., before latency saturates.
Note the logscale axis, to better capture  performance differences.
We increase the system size in increments of $6$, starting from the smallest size of $4$, until we reach $100$.

As an overall observation, our two \sysname prototypes outperform the consensus-based solution at every system size we investigate.
At small size, all systems exhibit their respective highest throughput.
The consensus-based implementation using BFT-SMaRt sustains over $10K$ \emph{pps}, while \sysname reaches almost $13.5K$ \emph{pps} and \sysnamesig sustains $55K$ \emph{pps}.
The $4$x improvement of \sysnamesig over \sysnameecho is owed to the the linear communication complexity of the former system (\Cref{sec:variants-details}).
As can be seen, however, this benefit slowly tapers off with increasing system size.
At maximum system size ($N=100$), the consensus-based system saturates at $334$ \emph{pps}; \sysnameecho sustains $6$x higher throughput, being able to apply $2K$ \emph{pps}, and \sysnamesig can sustain $5K$ \emph{pps} (a $16$x improvement over consensus and $2.5$x over \sysnameecho).

\begin{figure}
\centering
\includegraphics[width=\columnwidth]{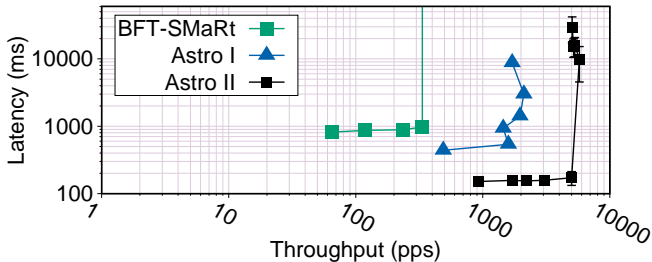}
\vspace{-14pt}
\caption{\textbf{Latency/throughput.} Performance evaluation of three payment systems each running at $N=100$.}
\vspace{-.4cm}
\label{fig:100-regional}
\end{figure}




\vspace{.1cm}\noindent\textbf{Latency-Throughput.}
We now explore the difference in performance between the consensus-based baseline and \sysname I/II at the maximum system size we consider, $N=100$.
As before, all systems are running in a single-shard setup.
The results depicted in~\Cref{fig:100-regional} show how latency evolves with respect to throughput.
For clarity sake, the y-axis (latency) starts at $100ms$, and we convey order of magnitude differences using logscale axes.

The consensus-based implementation typically exhibits sub-second latencies.
We do not show the $95^{th}$ percentile latencies because they obstruct visibility, but these are between $1.3$ and $1.5$ seconds.
Latencies in \sysnameecho are more variable, between $400$ and $500ms$ prior to saturation, while the $95{th}$ percentile latencies are on the order of one second.
Recall that clients connect to random replicas, which are geographically spread.
\sysnamesig exhibits more stable performance and lower latencies: prior to saturation, clients observe a confirmation latency of $200ms$ on average.
The $95^{th}$ percentile latency (at low load) is under $240ms$.
The $99^{th}$ percentile for all these systems are within the same order of magnitude as the $95^{th}$.

We remark that the latencies for these three systems are not necessarily at their worst when $N=100$.
We also investigate the same execution at $N=10$, for instance, and observe only slightly better performance (e.g., latency for \sysnamesig is $150ms$ on average).
The latencies do not change considerably because there is a lot of parallelism inherent in the underlying quorum-based protocols, both for consensus and broadcast.
This is intuitive: obtaining one response from a particular distant replica takes roughly as much time as obtaining several responses (in parallel) from multiple distant replicas.
Primarily, it is throughput that suffers in quorum-based systems, and latency secondarily~\cite{clement09making,van04chain,vuko15quest}.

An important observation here is that our evaluation concerns the critical part of a payment system, the ordering layer.
For the deterministic system model, we are only aware of prior experiments of this layer which considered a maximum system size of $N=10$, concretely for Hyperledger Fabric~\cite{sou18byzantine}, which builds on BFT-SMaRt.
To conclude this part of our evaluation, for systems of moderate size---up to $100$ replicas---broadcast-based systems are simpler and significantly outperform consensus-based solutions for decentralized payments.
Even if \sysname relies on broadcast, it still employs quorum-gathering to achieve consistent replication; hence the throughput of \sysname is inversely proportional to the system size (akin to consensus-based solutions).
To avoid this throughput decay and scale to larger systems, we now discuss experiments with sharding.

\subsubsection{Sharding in Smallbank Application}
\label{sec:smallbank-eval}

For a real-world application workload, we use the Smallbank transaction family from the {\small \uppercase{blockbench}} framework~\cite{dinh17blockbench}; this is a version of the H-Store Smallbank benchmark~\cite{cah09ssi} adapted to the cryptocurrency setting.
The application models bank accounts, where the owners of these accounts are clients that can issue several types of transactions. In particular, accounts can be of either \emph{savings} or \emph{checking} type. Some transactions model payments across two accounts of the same owner, while other transactions deal with the transfer of funds between different owners. 
For the sake of consistency, hereinafter we refer to bank \emph{accounts} and their \emph{owners} as \lognameplural and \emph{clients}, respectively.

\vspace{.1cm}\noindent\textbf{Experimental Setup.}
We associate each client with two \lognameplural (for checking and savings).
Thus same-client transactions at the application level appear as full-fledged payments between two distinct \lognameplural in the underlying layer.
We use a multi-shard setup for \sysnamesig, ensuring that both \lognameplural of any client belong to the same shard.
Whenever a transaction involves different shards, the cross-shard coordination consists of the \opname{Credit} message described earlier (\Cref{sec:sharding-mechanism}).
For BFT-SMaRt, we use an equivalent setup.

Each shard consists of $N=52$ replicas uniformly spread among the four EC2 regions of Europe.
We execute using $2, 3$ and $4$ shards (total of $208$ replicas); we limit ourselves to $4$ mainly due to financial constraints, but also because it is straightforward to estimate performance at larger scales.
Clients attach to a certain replica and simultaneously issue transactions as prescribed by the Smallbank benchmark, meaning that 12.5\% of the overall number of transactions are cross-shard.
To produce more realistic network conditions, we introduce artificial network delays:
We use the traffic control (\texttt{tc}) subsystem of the Linux Kernel, and increase inter-replica  latencies by $20ms$.
Network latency between replicas in Europe is around $20ms$, so having this delay essentially doubles latencies; additionally, this also eliminates any advantage that may arise due to co-location of some replicas in the same EC2 region.

\begin{table}[t]
\setlength{\tabcolsep}{0.5em}
\footnotesize
\begin{tabular}{c|c|c|c|c|c|c|}
\cline{2-7}
\multicolumn{1}{c|}{}
& \multicolumn{1}{|c|}{\centering \#}
& \multicolumn{1}{|p{1.3cm}|}{\centering \texttt{tc} \\ delay (ms)}
& \multicolumn{2}{|p{2.9cm}|}{\centering Throughput (Kilo-\emph{pps}) \\ per-shard$\setminus$\textbf{total}}
& \multicolumn{2}{|p{2.3cm}|}{\centering Latency (ms) \\ Average$\setminus\mathbf{95^{th}~\%ile}$} \\

\multicolumn{1}{c|}{}
& \multicolumn{1}{|c|}{}
& \multicolumn{1}{|c|}{}
& \multicolumn{1}{|c|}{\centering \sysnamesig}
& \multicolumn{1}{|c|}{\centering BFT-S$^{\dagger}$}
& \multicolumn{1}{|c|}{\centering \sysnamesig}
& \multicolumn{1}{|c|}{\centering BFT-S$^{\dagger}$} \\

\cline{2-7}
\multirow{6}{*}{\rotatebox[origin=c]{90}{\# of shards}}
& 2  & 0  & 7.9$\setminus$\textbf{15.7} & 1.0$\setminus$\textbf{2.0} & 204$\setminus$\textbf{279} & 600$\setminus$\textbf{808} \\
& 2  & 20 & 5.1$\setminus$\textbf{10.2} & 0.3$\setminus$\textbf{0.5} & 479$\setminus$\textbf{705} & 2245$\setminus$\textbf{2673} \\ \cdashline{2-7}
& 3  & 0  & 5.1$\setminus$\textbf{15.4} & 1.0$\setminus$\textbf{3.1}  & 213$\setminus$\textbf{375} & 600$\setminus$\textbf{808} \\
& 3  & 20 & 4.5$\setminus$\textbf{13.6} & 0.3$\setminus$\textbf{0.8}  & 368$\setminus$\textbf{656} & 2245$\setminus$\textbf{2673} \\ \cdashline{2-7}
& 4  & 0  & 5.0$\setminus$\textbf{20.1} & 1.0$\setminus$\textbf{4.1}  & 213$\setminus$\textbf{259} & 600$\setminus$\textbf{808} \\
& 4  & 20 & 4.5$\setminus$\textbf{18.1} & 0.3$\setminus$\textbf{1.1}  & 354$\setminus$\textbf{620} & 2245$\setminus$\textbf{2673} \\ \cline{2-7}
\end{tabular}

\caption{\textbf{Smallbank sharded benchmark.} Performance results for up to $4$ shards (each $N=52$ replicas). $^\dagger$BFT-SMaRt results are upper-bound values based on a single-shard experiment.}
\label{tbl:smallbank}
\vspace{-.35cm}
\end{table}

\vspace{.1cm}\noindent\textbf{Experimental Results.}
We provide the results in~\Cref{tbl:smallbank}.
We show both per-shard and overall (i.e., total) throughput for a given latency envelope.
\sysnamesig sustains the highest per-shard throughput when there are 2 shards.
As the number of shards increases (the \# column), per-shard throughput slowly decreases: This is because intra-shard payments are more lightweight (lacking the cross-shard notification mechanism) and the number of intra-shard operations decreases with growing number of shards~\cite{zama18rapidchain}.
We observe that the $20ms$ network delay affects performance.
The reason is TCP's congestion control: \sysnamesig saturates the links and network delays become the bottleneck.

As~\Cref{tbl:smallbank} shows, performance in \sysnamesig scales well with the number of shards.
In absolute numbers, \sysnamesig sustains up to $20K$ \emph{pps} using four shards, with average latencies of around $200ms$.
The BFT-SMaRt baseline running on four shards sustains a total throughput of just above $4K$ \emph{pps};
importantly, these result are only for comparison, and represent optimistic upper-bounds.
In BFT-SMaRt we omit the cross-shard coordination step, which typically consists of a 2PC protocol posing significant overhead, thus a fully working sharded solution would necessarily sustain less than $4K$ \emph{pps}~\cite{kok18omniledger,zama18rapidchain}.

\vspace{-1pt}
\subsection{Performance Robustness}
\label{sec:results-robust}

We now investigate how our \sysname and the baseline react to two problems that can arise in practice, namely \emph{failure} (e.g., crash) and \emph{asynchrony} (network delays) at a replica.
We consider the impact of these issues when they affect a random replica in each system, as well as the case when the leader is affected in the consensus-based system.

\sysnameecho and \sysnamesig have similar robustness characteristics: they are completely decentralized (there is no leader) and making a payment only requires broadcasting a message.
To maximize fairness of comparison, we experiment with \sysnameecho, as its message pattern and cryptographic primitives (MAC-based channel authentication) are the most similar to BFT SMaRt.


We study the evolution of throughput within a window of execution of $40s$, ignoring a warm-up period of $20s$.
For all these experiments, we introduce asynchrony or failure after $30s$ elapse.
To induce asynchrony, we again use the traffic control utility \texttt{tc} with the network emulator queuing discipline.
We always use a delay of $100ms$.
For instance, to introduce such a delay on all packets outgoing from interface \texttt{eth0} at a replica, we use the following command:

\texttt{\small tc qdisc change dev eth0 root netem delay 100ms}.

\begin{figure}[t]
\vspace{-.5cm}
\centerline{\includegraphics[width=0.40\textwidth]{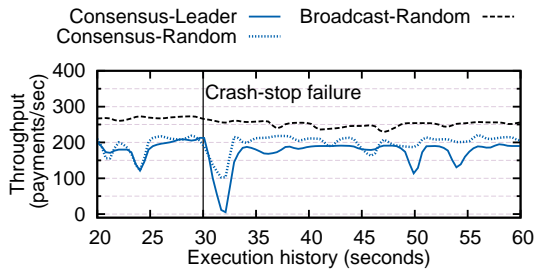}}
\vspace{-1pt}
\caption{\textbf{Throughput robustness during crash-stop failures.} We plot throughput when a replica crashes in the consensus-based system (either the leader or a random replica) and \sysnameecho .}
\vspace{-.35cm}
\label{fig:robust-fail}
\end{figure}

We use $10$ clients, each running a single thread.
The goal is to evaluate these systems below saturation point.
If we introduce failures at saturation, this can lead BFT-SMaRt to halt or enter a livelock where the system is unable to do view-change (i.e. leader election).
Moreover, at saturation point \sysname can sustain the same throughput independently of how many replicas accept client operations; this is because no single replica in our broadcast-based system is a bottleneck.
In other words, stopping a replica at saturation point in \sysname would not impact throughput, giving an advantage to our system over the consensus-based solution.
We first report results for a system size $N=49$.
We run these experiments with larger and smaller systems, but similar observations emerge as the ones we describe below.
For completeness, we also discuss a set of interesting results with a larger system size of $N=100$.

\begin{figure}[t]
\vspace{-.5cm}
\centerline{\includegraphics[width=0.40\textwidth]{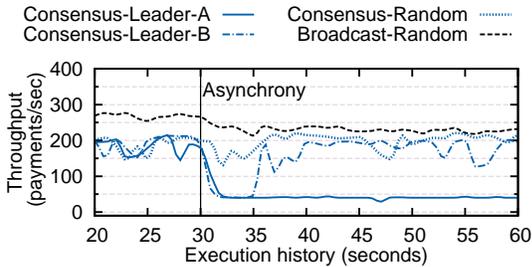}}
\vspace{-1pt}
\caption{\textbf{Throughput robustness during asynchrony.} We show, for N=49, how throughput evolves in the consensus- and broadcast-based systems during asynchrony ($100ms$ delay for each outgoing packet) at one replica, either the leader or random.}
\vspace{-.35cm}
\label{fig:robust-async}
\end{figure}

In~\Cref{fig:robust-fail} we show how the throughput evolves when we introduce a crash-stop failure at a replica ($N=49$).
For consensus, this failure has a severe impact on throughput if the leader is affected (the \emph{Consensus-Leader} curve), because the view-change protocol has to execute.
The throughput drops to $0$ while this protocol runs, typically a few seconds.
For larger system sizes, this protocol can take longer to execute, as we will show later.
When a random replica fails in the consensus-based system (\emph{Consensus-Random}), there is a brief decrease in throughput when all clients and replicas get disconnected from the affected replica, but thereafter performance recovers.
In \sysnameecho we stop a random replica (\emph{Broadcast-Random}), and thereafter throughput drops from $~270$ pps to $~250$ pps, which accounts for the failed replica which was handling roughly $20$ pps from one of the clients.
This decrease is barely visible in the plots.

\Cref{fig:robust-async} shows how asynchrony impacts the performance in the two systems ($N=49$).
We depict two separate executions for the case of consensus when the leader is affected, because there are two possible outcomes.
First, it may happen that throughput decreases and remains that way; this is the \mbox{\emph{Consensus-Leader-A}} timeline.
Second, the system can go through a view-change (\mbox{\emph{Consensus-Leader-B}}) because the leader is too slow or its buffers can overflow and packets get dropped (inflating the replica-to-replica delay).
Clearly, initiating a view-change is preferable in this case, because the throughput penalty is smaller.
There is a well-known tradeoff, however, in choosing the view-change timeout~\cite{clement09making,mil16honeybadger}: initiating view-change too aggressively can lead to frequent leader changes even in good conditions, which can erode performance on the long-run.

When a random replica is affected with asynchrony in the consensus-based system (\emph{Consensus-Random} execution in~\Cref{fig:robust-async}), performance drops briefly because there is a quorum switch, i.e., the affected replica is replaced by a different one in the active quorum~\cite{antoni18smr}.
For the broadcast-based system (the \emph{Broadcast-Random} timeline), asynchrony affects performance in the same manner in which a failure does.
Concretely, the affected replica no longer sustains the same amount of client operations, so the overall throughput reduces correspondingly.

\begin{figure}[t]
\vspace{-.5cm}
\centerline{\includegraphics[width=0.40\textwidth]{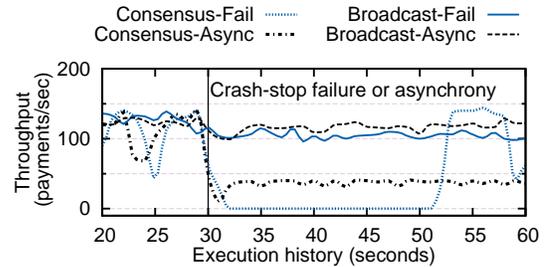}}
\vspace{-1pt}
\caption{\textbf{Throughput robustness.} We show how throughput evolves for $N=100$ when a crash-stop failure or asynchrony affects the consensus-based system or the broadcast-based system.}
\vspace{-.35cm}
\label{fig:robust-100}
\end{figure}

We also show results for the case of a larger system size ($N=100$) in~\Cref{fig:robust-100}.
There are four timelines in this execution, as follows.
For the consensus-based solution, we show what happens when there is either a  crash-stop failure or asynchrony at the leader.
In the former case (\emph{Consensus-Fail}), the view-change protocol kicks off and lasts for roughly $20$ seconds, while throughput stays at zero; this is similar to the \emph{Consensus-Leader} execution in~\Cref{fig:robust-fail}.
In the latter case (\emph{Consensus-Async}), performance degrades and stays that way for as long as the affected replica remains the leader; this is similar to the \emph{Consensus-Leader-A} in~\Cref{fig:robust-async}.
For the broadcast-based solution we consider the same two issues affecting a random replica.
When either of these issues arises (\emph{Broadcast-Fail} or \emph{Broadcast-Async}) throughput is affected correspondingly with the number of operations that the failed replica is handling (and which is unable to continue).
Note that \sysname relies on fate-sharing~\cite{cl88design} between a client and its representative: when a replica stops, all the associated \lognameplural naturally stop as well.

We conclude with two general observations.
First, \sysname does not suffer from overall (i.e., global) throughput degradation that can happen in leader-based protocols such as most consensus algorithms.
Second, our system does not rely on timeouts for liveness.
Simply put, \sysname progresses at the speed of the network.
These two advantages are closely linked, and they both follow from the asynchronous nature of the broadcast protocol we rely on.



\vspace{-3pt}
\section{Related Work}
\label{sec:related}

Since Nakamoto's original Bitcoin paper~\cite{nakamotobitcoin}, follow-up payment systems seek to prevent double-spending by establishing a total order of transactions, i.e., solving consensus.
Consequently, a lot of effort has been devoted to improving the consensus layer.

\vspace{.1cm}\noindent\textbf{Dealing with the Consensus Bottleneck.}
Research on consensus algorithms has shown significant breakthroughs and modern protocols quote impressive performance numbers~\cite{Yu18RepuCoin,zama18rapidchain}.
To push performance even further, several interesting systems address the consensus bottleneck with sharding~\cite{al2017chainspace,kok18omniledger,luu2016secure,zama18rapidchain,wang2019monoxide}. 
Approving a cross-shard payment, however, requires special coordination~\cite{kok18omniledger,wang2019monoxide,zama18rapidchain}.
Off-chain payment networks such as 
Lightning~\cite{po16bitcoin} and Raiden~\cite{raiden}
strive to minimize the impact of consensus protocols.
They allow parties to move funds from a blockchain into high-performance payment \emph{channels}, for which the final balance is settled back on the blockchain after use.
Recent advances in this field rely on trusted hardware to provide an asynchronous protocol for all interactions~\cite{lind2017teechain}.
These results bring noticeable improvements over Bitcoin, enabling good scalability and very fast payments.
Nevertheless, the underlying problem of consensus is only reduced, not overcome.
In \sysname we take a different approach: We provide robust performance by avoiding consensus protocols altogether.




\vspace{.08cm}\noindent\textbf{Performance Instability of Consensus.}
Recent work emphasizes the problem that performance of consensus 
algorithms hangs on a fragile thread, namely their view-change sub-protocol~\cite{buchman2018latest,guerraoui2019100,yin19hostuff}.
HotStuff, for instance, proposes to absorb view-change in the common-case consensus algorithm;
this sidesteps performance instability but comes with the cost of a higher common-case latency~\cite{yin19hostuff}.

Another line of research circumvents the view-change issue with randomized consensus protocols, such as HoneyBadgerBFT~\cite{mil16honeybadger} or BEAT~\cite{du18beat}.
Both are based on work by Ben-Or et al.~\cite{ben1994asynchronous} combining  reliable broadcast (\brbroadcast) with  binary Byzantine agreement (\textsc{aba}).
In a nutshell, these protocols comprise a broadcast phase (where replicas form encrypted batches of payments which they disseminate using \brbroadcast), an agreement phase (involving $N$ instances of the \textsc{aba} protocol to agree on a common set of batches), and a decryption phase (requiring each replica to obtain $f+1$ decryption shares).
These protocols push the performance of consensus by carefully choosing modern cryptographic tools and system parameters.

Various \emph{leaderless} consensus protocols have been proposed, for both crash and Byzantine models~\cite{schiperLeaderFree,cor05low,crain2018dbft,lamportLeaderless,moraru2013there}.
These protocols, however, either make use of some form of coordinator in corner-cases, or rely on additional synchrony assumptions, or provide probabilistic guarantees.
For example, a thorough study of the appendix of \cite{moraru2013there} reveals that EPaxos only ensures probabilistic liveness and, as shown recently \cite{epaxosBug}, has correctness issues.

\sysname is deterministic and fully asynchronous. It does not solve the general consensus problem, but instead focuses on payments.
Since our system relies exclusively on \brbroadcast and no \textsc{ba} primitive is necessary, \sysname is simpler and more efficient than modern leaderless randomized consensus protocols.

\vspace{.08cm}\noindent\textbf{Avoiding Consensus Protocols.}
Recent theoretical results~\cite{guer19cryptonumber,gup16nonconsensus}
show that consensus is unnecessary for implementing a payment system, contrary to popular belief. 
For instance, ~\cite{guer19cryptonumber} showed that the basic double-spending problem, as defined by Nakamoto~\cite{nakamotobitcoin}, can be cast as a sequential object type and that it has consensus number 1 in Herlihy's hierarchy~\cite{Her91}. 
Whilst the observation that consensus is unnecessary to prevent double-spending in a theoretical context has been made, we apply this insight for the first time to obtain \sysname: a full system solution (design, implementation, evaluation), that is also efficient.

The \lognamelongplural in \sysname resemble conflict-free replicated data types (CRDTs)~\cite{sh11crdt}.
Similar to a CRDT, different \lognameplural support concurrent updates while preserving consistency.
Since each log has a unique owner, we rule out the possibility of conflicting operations on each log.
Note, however, that appending a payment to the history of a client's \logname \emph{A} is not commutative, i.e., any two payments within the same history need to be ordered with respect to one another.  This is a departure from classic CRDTs, but it ensures in our case that the state at correct nodes always converges to a consistent version.

Our \logname abstraction in \sysname resembles 
the acyclic graph (DAG) in various novel payment systems~\cite{ chu6byteball,he16corda,kar18vegvisir,som13accelerating}.
The distinctive feature of \sysname, however, is that consensus is entirely sidestepped---whereas all prior solutions we are aware of, even those building on a DAG, employ a consensus algorithm to order payments.

%


\vspace{.08cm}\noindent\textbf{Broadcast Protocols.}
\brbroadcast protocols have a long tradition starting with the algorithms of Bracha and Toueg~\cite{bra87asynchronous,br85acb}.
Later work refined and improved performance and properties of these algorithms~\cite{cach02sintra,ma97secure,MR97srm,reit94}.
Asynchronous verifiable information dispersal algorithms~\cite{cachin2005asynchronous} are closely related to \brbroadcast protocols, and both of these classes of protocol represent an essential building block in modern asynchronous consensus protocols~\cite{du18beat,mil16honeybadger}.

There are several ways to improve the scalability of broadcast protocols.
Sharding---the technique we recalled out above and we adopt in \sysnamesig---is a clean approach to scalability, as it allows each shard to maintain the same (deterministic) properties as a non-sharded system.
Other approaches, such as clustering~\cite{guer16atum,scheideler2005spread}, probabilistic quorum-based~\cite{malk01proba}, or sample-based~\cite{guer19ssb},
typically yield a design providing probabilistic guarantees.

\section{Conclusions}
\label{sec:conclusion}
\sysname is a decentralized payment system that can sustain $20K$ payments/sec in a deployment of 200 replicas, while exhibiting sub-second latency.
It can do so by not relying on any consensus layer and thus remaining mostly unaffected by network asynchrony and compromised replicas.
We do not claim \sysname to be a silver bullet: we only focused on payments and did not consider the general abstraction of state machine replication, e.g., as might be required by smart contracts. Yet, determining the exact set of problems (besides payments) that can be addressed by \sysname's broadcast layer is an open problem. We also identified several avenues for improving \sysname, namely: \emph{(1)} a more flexible representation scheme, instead of the fixed dependency between a client and its representative replica, \emph{(2)} use more advanced cryptographic primitives (e.g., threshold signatures, key revocation schemes),
\emph{(3)} a fine-grained state transfer protocol for reconfiguration,
and \emph{(4)} a hybrid system that incorporates asynchronous payments and consensus-based smart contracts.

\newpage

\bibliographystyle{acm}
\bibliography{refs}

\newpage \appendices

\section{Asynchronous Reconfiguration}
\label{sec:reconf-appendix}

The description of \sysname is focused on a static system with a fixed set of replicas and clients, in order to clearly present its design.
For long-lived systems, which we expect a payment system to be, adding and removing replicas is desirable, e.g., if participants decide to start/stop using the payment system, or when replacing old replica machines by new ones.

Reconfiguration of clients is straightforward: each client has a representative replica \emph{r}, so adding a client simply means that \emph{r} executes a \brbroadcast instance announcing the new client; subsequently all replicas start maintaining the \logname of this new client.
Reconfiguration of system replicas is a more challenging problem, which we discuss in the rest of this section.

In consensus-based systems, reconfiguration can be handled by the consensus module.
For instance, BFT-SMaRt~\cite{bessani2013efficiency} and similar systems~\cite{lam09vertical} treat a reconfiguration request as a special request which is totally-ordered just like ordinary client requests.

Consensus, however, is not always necessary for reconfiguration.
For example, DynaStore~\cite{agu10reconfig} and FreeStore~\cite{alc16efficient} provide solutions for consensusless reconfiguration of read/write storage in the asynchronous crash-stop model.


The purpose of this appendix is to briefly present a line of research that -- we believe -- answers in the affirmative the question of whether reconfiguration is possible for a payment system in the Byzantine model. 
The consequence is that our payment system does not require consensus throughout the entirety of its lifetime, which eradicates any possible argument supporting the necessity of consensus.
In this line of research, we adopt ideas from the FreeStore~\cite{alc16efficient} protocol, and to account for the Byzantine failure model we build on Byzantine quorum systems~\cite{malk97byzantinequorums}.
Admittedly, the details of reconfiguration are non-trivial and a thorough explanation of it is an independent publication. Our goal is to present a high-level overview of our ongoing result.

\subsection{Overview}
\label{sec:reconf_overview}
Throughout the lifetime of a system, each correct replica passes through a sequence of numbered \emph{views}. A view is a set of replicas that a replica considers to constitute the system. At any point in time, each replica has exactly one \emph{current view}.

The interface of the reconfiguration protocol exposes operations \emph{Join}/\emph{Leave}. Those operations consist of broadcasting a \opname{Join}/\opname{Leave} message to some view $v$, which represents the current state of the system as seen from the perspective of the joining/leaving replica.

Our reconfiguration protocol guarantees that, for a finite number of reconfiguration requests in any execution, all replicas converge to a single final view which incorporates every reconfiguration request issued in the execution.
We say that a view \emph{v} is \emph{installed} if some correct replica considers \emph{v} its current view and processes payment operations in \emph{v}. 
Moreover, our reconfiguration protocol ensures that the installed views form a sequence.
Our state transfer protocol simply consists of sending all \lognameplural to the joining replica.

We adapt the payment protocol so that all messages include the current view of the sending replica.
Correct replicas behave consistently across views with respect to each payment.

When a replica \emph{r} observes a view that is more recent than \emph{r}'s current view, \emph{r} pauses payment execution.
Roughly speaking, \emph{r} resumes execution after coordinating with a quorum of replicas belonging to the new view, and then \emph{r} executes payments assuming the membership of the new view.
Since reconfiguration is not a very frequent operation, we expect the overall downtime caused by reconfiguration to be insignificant.
We evaluate the reconfiguration overhead of joining replicas in the next section.



\subsection{Evaluation}
\label{sec:reconf-eval}
The experiment of asynchronous reconfiguration starts with a system of $N=4$ replicas; subsequently, new replicas join the system until $N=80$, one by one.
Note that our reconfiguration protocol allows batched joins (which we avoid so that we can measure the latency of the protocol itself), and that all replicas are randomly distributed across Europe (\Cref{sec:methodology}).
During this experiment, the system is quiescent, i.e. no client submits any payment.

\begin{figure}[t]
\centerline{\includegraphics[width=\columnwidth]{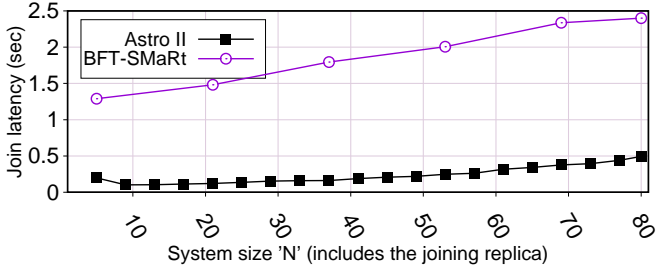}}
\caption{\textbf{Reconfiguration latency}. Latency (seconds) of a join operation at different system sizes, in \sysnamesig and in BFT-SMaRt.}
\vspace{-.3cm}
\label{fig:reconf}
\end{figure}

We measure the latency (i.e. time to join) both for \sysnamesig and BFT-SMaRt in \Cref{fig:reconf}.
In \sysnamesig , latency represents the elapsed time between the moment when the joining replica sends the reconfiguration request until this replica becomes able to participate in the payment protocol.
Latency in BFT-SMaRt represents the elapsed time between sending the special type of operation by the View Manager \cite{bess14state} and sending message to the joining replica that it can start participating in the protocol and should get up-to-date with the rest of system. 
The first data point for \sysnamesig shows slightly higher latency than for subsequent points, which is due to the fixed overhead of establishing connections between replicas already in the system. 
As~\Cref{fig:reconf} shows, latency in BFT-SMaRt is an order of magnitude higher than in \sysnamesig.
We are not aware of any
published numbers on consensus-based reconfiguration latency; we believe that the primary reason for this difference in performance is owed to a simpler, more efficient protocol.

\subsection{Dynamic Byzantine Reliable Broadcast}
Dynamic Byzantine Reliable Broadcast (\dbrbroadcast) represents the continuation of the work briefly introduced earlier (\Cref{sec:reconf_overview}).
An in-depth theoretical analysis of \dbrbroadcast, along with a thorough proof of its correctness, is provided in \cite{guerraoui2020dynamic}.

Instead of using \brbroadcast based on Bracha's algorithm, \sysnameecho can adopt \dbrbroadcast as an underlying broadcast layer.
Since \dbrbroadcast provides the exact same properties (adapted to the dynamic environment) as those of the broadcast layer (including totality), no further modifications of \sysnameecho are needed.
Given that \sysnamesig does not require the broadcast layer to provide totality, we can use a modified version of \dbrbroadcast that does not provide totality (denoted with \dbrbroadcasttotality).

The algorithm for \dbrbroadcasttotality is obtained from the \dbrbroadcast by excluding the last ``all-to-all'' ~\cite[\text{Section 4.3}]{guerraoui2020dynamic} communication step. With this modification, \dbrbroadcasttotality becomes a direct replacement of the \brbroadcast within \sysnamesig.


\comment{
A \emph{view} is a fundamental concept in reconfiguration. A view $v$ is composed of a set of $updates$, where update $u$ states whether some replica $r$ wants to join or leave the system. Hence, $v$ specifies the set of replicas that are members of the system at a certain point in time. Due to asynchrony, each replica may have a different perception of the set of replicas that belong to the system, i.e. it may have a different \emph{current view}. A view $v$ is \emph{installed} if some correct replica considers $v$ its current view and that replica processes payments in $v$.

Our reconfiguration protocol comprises three phases:
\begin{enumerate}
  \item \textit{Negotiation}: During this phase, a (correct) replica that wants to join or leave the system learns the most updated view installed in the system.
  \item \textit{Convergence}: This phase ensures that the installed views, through which the system transits over time, form a totally ordered sequence $s$, such that the following stands:
  \begin{equation*}
  \forall v_i, v_j: v_i \in s \land v_j \in s \land i < j \implies v_i \subset v_j
  \end{equation*}
  \item \textit{View Installment}: During this phase, a (correct) replica updates its current view and the replicas that are joining get up-to-date with the rest of the system.
\end{enumerate}

For the rest of this appendix, we detail these three phases (\Cref{app:reconf-Negotiation,app:reconf-Convergence,app:viewchange}) and a few additional clarifications (\Cref{app:additional-details}).
We will present the actions of a replica that wants to issue a reconfiguration request (i.e. join or leave), as well as the way the rest of the system handles this request.
Before that, however, we will state our assumptions and safety/liveness properties.

\paragraph{Assumptions.}
First, we assume that number of reconfiguration requests is finite.
Second, we build on the basic upper bound for Byzantine faults of $N/3$, which applies to any view. 
Hence, we assume that the initial view respects the mentioned bound for Byzantine faults.
Lastly,



\paragraph{Safety Property.}
In the basic payment protocol (\cref{sec:payment-protocol}) double-spending is prevented by the underlying \brbroadcast of \sysname.
Nevertheless, with the introduction of reconfiguration double-spending may happen among different views of the system, unless we explicitly prohibit it. To this end, our reconfiguration protocol satisfies the following safety property: If a payment $a$ with identifier $(s, n)$ is settled in a view $v_1$, then no other payment $b$ with the same identifier $(s, n)$ can be settled in a view $v_2 \neq v_1$.

\paragraph{Liveness Properties.}
Briefly speaking, liveness properties ensure that in \sysname no payment broadcast by an honest replica hangs forevers, even when a payment is issued during system reconfiguration. More precisely, (1) a (correct) replica $r$ that wants to join/leave the system, eventually does so, (2) every payment broadcast by an honest replica eventually settles (i.e. there exists a view $v$ in which the payment settles).


\subsection{Negotiation Phase}
\label{app:reconf-Negotiation}
Let $r$ be a correct replica that wants to join (resp. leave) the system.
Replica $r$ has its own current view $cv_r$ of the system membership. 
At this point, $r$ broadcasts the message $m$ = $\left<\textrm{\opname{ReconfigReq}}, \textrm{\opname{Join}}, r, cv_r\right>$ (resp. $\left<\textrm{\opname{ReconfigReq}}, \textrm{\opname{Leave}}, r, cv_r\right>$).

Let $q$ be a correct replica whose current view is $cv_q$, and which receives $m$. Replica $q$ will acknowledge $m$, by sending $\left<\textrm{\opname{ReconfigAck}}, cv_q\right>$, if $cv_q = cv_r$. 
\needsrev{However, if $cv_q \supset cv_r$ then replica $q$ replies to $r$ with the message $\left<\textrm{\opname{Amend}}, cv_q\right>$, indicating that $r$ should re-broadcast the reconfiguration request to view $cv_q$.}

The convergence and the view installment phase of our protocol ensure that if the replica $r$ receives $2f_{cv_r}+1$ $\left<\textrm{\opname{ReconfigAck}}, cv_r\right>$ messages, then the request of  $r$ will be processed in the next instance of reconfiguration. 
\needsrev{If, instead, $r$ receives message $\left<\textrm{\opname{Amend}}, v', proof\right>$ and $cv_r \subset v'$, then $r$ will adopt $v'$ as its current view (i.e. $cv_r = v'$) and re-broadcast the join (resp. leave) message in $v'$.}

\needsrev{Note that a malicious process can create a view $v'$ that contains of more than a $1/3$ of Byzantine replicas. In this scenario, a joining replica is trapped in view $v'$ and is not able to participate in the payment protocol. We avoid this issue by using the $proof$ argument of the \opname{Amend} message. In particular, the sender of the \opname{Amend} message must provide a certificate that view $v'$ can be tracked back to the initial view.}

\subsection{Convergence Phase}
\label{app:reconf-Convergence}
The convergence phase of our reconfiguration protocol uses a \emph{view generator}, whose abstraction is discussed in detail in FreeStore~\cite{alc16efficient}. 
FreeStore's design assumes that failures are crash-stop. Our system, instead, assumes Byzantine failures. We generalize the view generator presented in FreeStore to the Byzantine setting. We achieve this by (1) building on a Byzantine reliable broadcast (\brbroadcast), and (2) adjusting the quorum system.
We describe each in more detail.

\paragraph{Byzantine Reliable Broadcast (\brbroadcast).} 
We replace best-effort broadcast within the view generator in FreeStore with a \brbroadcast building block featuring strong reliability (\cref{appendix:brb-echo}).
Note that it is of the crucial importance that the broadcast primitive ensures string reliability property.
\brbroadcast is based on Bracha's algorithm~\cite{br85acb} and is described in detail in \Cref{appendix:brb-echo}.

The internal operation of the view generator involves the exchange of messages in a manner predefined by the view generator protocol. A correct replica $r$ will deliver a message $m$ only when $r$ has already delivered the predefined set of messages $D$. We say that messages in $D$ are the \emph{dependencies} of message $m$. 
 
\needsrev{The correctness of the view generator--as described in FreeStore--breaks under the Byzantine failure model. 
For example, the termination property of the view generator (in FreeStore) relies on replicas down-calling it at most once per view, but in a Byzantine environment malicious replicas may not adhere to this. 
We combined \brbroadcast along with the aforementioned concept of dependencies in order to enable every correct replica $r$ to easily determine whether some message $m$ (received, but not yet delivered by $r$) should have been sent or not.
Replica $r$ ignores $m$ if it should not have been sent.}
 
\paragraph{Quorum System.} The view generator tied with view $v$ of FreeStore employs quorums of size $f_v+1$ (for a view size of $2f_v+1$), which is common for protocols operating under the crash-stop failure model. 
Replacing best-effort broadcast with \brbroadcast---as we mentioned above---is not enough by itself. The issue is that \emph{join/leave liveness} can be violated. Below we explain why having quorums of $2f_v+1$ replicas (for a view size of $3f_v+1$) solves this issue. \TODO{rewrite this}
 
Let $r$ be some (correct) replica that joins the system. 
Prior to joining, $r$ broadcasts the message $\left<\textrm{\opname{ReconfigReq}}, \textrm{\opname{Join}}, r, cv_r\right>$ and collects $2f_{cv_r}+1$ matching $\left<\textrm{\opname{ReconfigAck}}, cv_r\right>$ messages. This implies that at least $f_{cv_r}+1$ correct replicas, that are already part of the system in view $cv_r$, are aware of the reconfiguration request made by replica  $r$. 

To ensure join/leave liveness, it suffices to show that the request of replica $r$ will be processed in the next instance of reconfiguration. 
Indeed, by the intersection property of quorums and the design of the view generator protocol the following holds: 
if $v$ does not include a reconfiguration request (i.e., the \opname{ReconfigReq} message), then it is impossible to collect a quorum of proposals for a view $v$.

The convergence phase of the reconfiguration protocol for a correct replica is always guaranteed to finish. When a correct replica receives an up-call from the instance of the view generator that is tied to a view $v$, the replica uses the content of the up-call to update its view to $v'$. 
Using a \brbroadcast implementation with the strong reliability property (described in detail in~\Cref{appendix:brb-echo}, but adapted to generic messages instead of payments) ensures that every correct replica will eventually receive the up-call and make the transition to $v'$.


\subsection{View Installment Phase}
\label{app:viewchange}

A correct replica $r$ enters this phase upon receiving an up-call from the view generator.
In this phase, this replica will stop participating in the payment execution protocol (described in~\Cref{sec:payment-protocol}).
This phase comprises, in the common-case, two broad actions on the part of $r$, assuming that $r$ is part of the old view (the view that is being currently replaced), namely: (1) helping all the replicas get up-to-date (i.e., state transfer), and (2) helping the leaving replica $z$ to leave the system. We explain all of these in more detail below.

\paragraph{State Transfer.}
Replica $r$ helps replica $q$, that is a member of the view that is being installed (note that this replica may be joining replica), learn the system state by essentially transferring all the state which $r$ has, including any payment that $r$ has acknowledged but not yet delivered.
This is important to prevent double-spending across multiple views, which we elaborate upon later (\Cref{app:additional-details}).
The replica $q$ learns about the system state from at least $2f_{pv}+1$ replicas that were members of the previous view $pv$.
The \opname{StateUpdate} message has an additional purpose---to ensure that only one view $v$ can be considered the most up-to-date installed view in the system. 
More concretely, as soon as a replica sends this message, it will wait for $2f_{pv}+1$  \opname{StateUpdate} messages. 
Since at least $f_{pv}+1$ of received \opname{StateUpdate} messages have come from the correct replicas (which had previously stopped processing payment transactions), it is impossible to receive $2f_{pv}+1$ \opname{Ack} messages for the previous view $pv$. 
This ensures that only the view that is being installed can be considered the most up-to-date view installed in the system. 
Moreover, $r$ will only send \opname{Ack} messages corresponding to \opname{Prepare} messages that refer to this view (see the \sysnamesig description in~\cref{sec:variants-details} and~\Cref{appendix:brb-echo}).
At this point, replica $r$ also resumes participating in the payment execution protocol.

\paragraph{\opname{ViewUpdated} Message.}
Replica $r$ broadcasts a $\langle \textrm{\opname{ViewUpdated}}, cv_r \rangle$ to all the replicas that want to leave the system, and it does so after it updates its current view (i.e. $cv_r$ is the view that has just been installed). A leaving replica $z$ waits for $2f_v+1$ of mentioned messages, where $v$ is the view that is currently installed and exits the system.

We remark that some special cases can appear in this last phase of reconfiguration, particularly because of the asynchronous nature of the network.
For instance, a replica $r$ might transition across multiple views at once (i.e., from a very outdated view to another one); or the sequence of views has a length larger than one; or there is more than a single joining or leaving replica.
For these cases, we adopt the mechanisms of FreeStore~\cite{alc16efficient}.

\subsection{Additional Details}
\label{app:additional-details}

Reconfiguration introduces problems that do not exist if Astro system was static. This comes as no surprise as different replicas can have a different perception of the system membership. 
We now discuss such problems, as well as our approach for tackling them.

\paragraph{Reconfiguration Integration with Payment Protocol.}
Since different replicas can consider different replicas as current members of the system, each payment operation must be associated with a view. 
A (correct) replica $r$ that wants to issue a payment $a$ first broadcast message $m$ defined as $\left<\textrm{\opname{Prepare}}, a, cv_r\right>$ to view $cv_r$. 
When a correct replica $q$ delivers message $m$, $q$ must check whether its current view $cv_q$ is equal to $cv_r$. 
We need to analyse two different cases.
Notice that all three of \opname{Prepare}, \opname{Ack}, and \opname{Commit} in this description are message that belong to the \brbroadcast layer of the payment protocol (\Cref{sec:system-versions,appendix:brb-echo}).

\begin{enumerate}
    \raggedright
  \item $cv_q=cv_r$: In this case, replica $q$ sends a signed message $\left<\textrm{\opname{Ack}}, \mathrm{hash}(m), cv_q\right>$ to replica $r$.
  \item $cv_q \supset cv_r$: In this case, replica $q$ ignores message $m$. 
\end{enumerate}

In case replica $r$ receives $2f_{cv_r}+1$ \opname{Ack} messages (case 1),
then $r$ creates a certificate \emph{c} which comprises these acknowledgement messages and broadcasts $\left<\textrm{\opname{Commit}}, \mathrm{hash}(m), \emph{c}, cv_r\right>$ to view $cv_r$.
However, if $r$ receives $\left<\textrm{\opname{Amend}}, v' \supset cv_r, proof\right>$---where $proof$ proves that the view $v'$ is valid (i.e. it can be tracked down to initial view)---then $r$ waits to install some view $v'' \supseteq v'$ and broadcasts again $\left<\textrm{\opname{Prepare}}, a, v''\right>$ message to view $v''$. 
Notice that properties of the view generator ensure that replica $r$ eventually installs view $v''$.

\paragraph{Preventing Double-Spending.}
Since the system can be reconfigured, the problem of double spending across two views must be addressed.
Recall that our assumption in this protocol is that any two installed views intersect in at least $2f+1$ replicas.
Even with this assumption, it is possible that some payment $a$ with identifier \emph{(s, n)} can gather $2f_v+1$ acknowledgements in a certain view $v$, and then another payment $a'$ with the same identifier could gather $2f_{v'}+1$ acknowledgments in a subsequent view $v'$, leading to double spending.
To solve this issue, all replicas in view $v'$ communicate with at least $2f_v+1$ replicas from the previous view $v$ to find out which payments are unsettled in view $v$.

More precisely, three cases can appear when processing payments across views, and we now describe these cases alongside the respective approach we take to prevent double spending. 

\begin{compactenum}
    \item There exists only one payment $a$ for identifier \emph{(s,n)} and no other conflicting payment. In this case, replicas in view $v'$ are allowed to acknowledge $a$ in the new view $v'$.
    
    \item There exists at least one payment $a'$ conflicting with $a$. 
    In this case, replicas in view $v'$ would ignore any \opname{Prepare} message for any payment with identifier \emph{(s,n)}, since the presence of this conflicting payments represents effectively a proof-of-misbehavior for spender client $s$.
    
    \item There exists no payment whatsoever that had a \opname{Prepare} message with identifier \emph{(s,n)} in view $v$. In this case, replicas in view $v'$ can acknowledge any payment for this identifier.
\end{compactenum}
}
\section{Broadcast Layer Algorithms}
\label{appendix:brb-echo}

In~\Cref{lst:brb-echo} we sketch the algorithm implementing \brbroadcast based on the work of Bracha and Toueg~\cite{br85acb}.
We use this algorithm to build the broadcast layer in \sysnameecho.

\begin{lstlisting}[mathescape,
 caption={\brbroadcast protocol which we use in \sysnameecho , based on~\cite{br85acb}.},
 language=Astro,
  float,
 label={lst:brb-echo}]
// Called at replica 'r' to broadcast a message 'a'.
func Broadcast(a):
  prep := <PREPARE, a> (*@\label{line:send-phase}@*)
  sendToAll(prep) // Send to all replicas. (*@\label{line:sendtoall-1}@*)

// Process a protocol message m received from replica 'q' at 
// replica 'r'.
callback receive(q, m = <PREPARE, a>): 
// Handler for PREPARE messages.
  let a be {s, ts, _, _}
  if echoSent[q, (s, ts)] == false:
    echoSent[q, (s, ts)] := true
    sendToAll(<ECHO, q, (s, ts), a>)   (*@\label{line:sendtoall-2}@*)

callback receive(q, m = <ECHO, r, (s, ts), a>): 
// Handler for ECHO messages.
  echoes[r, (s, ts), a] += q
  if (|echoes[r, (s, ts), a]| (*@$\ge$@*) 2F+1) && (*@\label{line:sending-ready-1}@*)
      (readySent[r, (s, ts), a] == false):
    readySent[r, (s, ts)] := true
    sendToAll(<READY, r, (s, ts), a>)   (*@\label{line:sendtoall-3}@*)

callback receive(q, m = <READY, r, (s, ts), a>): 
// Handler for READY messages.
  readys[r, (s, ts), a] += q
  if (|readys[r, (s, ts), a]| (*@$\ge$@*) F+1) && (*@\label{line:sending-ready-2}@*)
        (readySent[r, (s, ts), a] == false):
    readySent[r, (s, ts), a] := true
    sendToAll(<READY, r, (s, ts), a>)   (*@\label{line:sendtoall-4}@*)
  if (|readys[r, (s, ts), a]| (*@$\ge$@*) 2F+1) &&
        (delivered[r, (s, ts), a] == false) &&
        (ts == allTS[s] + 1):
    delivered[r, (s, ts), a] := true
    trigger Deliver(a) (*@\label{line:fifo-deliver}@*)
    allTS[s] += 1
\end{lstlisting}
 
\sysnamesig uses a \brbroadcast implementation based on digital signatures~\cite{MR97srm}, which we detail in~\Cref{lst:brb-sig}.
\begin{lstlisting}[mathescape,
 caption={\brbroadcast protocol based on digital signatures, inspired by early work of Malkhi and Reiter~\cite{MR97srm}, which we use in \sysnamesig.},
 language=Astro,
 float, 
 label={lst:brb-sig}]
// Called at replica 'r' to broadcast a message 'a'.
func Broadcast(a)
  prep := <PREPARE, a>
  // Send the prepare message to all replicas.
  sendToAll(prep)

// Process a protocol message m received from replica 'q' at 
// replica 'r'.
callback receive(q, m)
  if (m = <PREPARE, a>)
    let a be {s, ts, _, _}
    pending[(s, ts)] := a
    sig := Sign(m)
    ackMsg := <ACK, (s, ts), sig>
    unicast(q, ackMsg) // Reply to replica q with ACK.
  else if (m = <ACK, (s, ts), sig>)
    return if invalidSignature(sig)
    acks[(s, ts)] := acks[(s, ts)] (*@$\cup$@*) {(q, sig)}
    if (|acks[(s, ts)]| == 2f+1)
      commitMsg = <COMMIT, (s, ts), acks[(s, ts)]>
      sendToAll(commitMsg) // Broadcast the commit message.
  else if (m = <COMMIT, (s, ts), proof>)
    return if (|proof| $<$ 2f+1) || (invalidSignatures(proof))
    a := pending[(s, ts)]
    // Release payment 'a' to the payment layer
    trigger Deliver(a)
\end{lstlisting}

For completeness, we also provide pseudocode describing the use of dependencies, i.e., the optimization that allows \sysnamesig to resolve the partial payments attack and enable sharding. 
To address these issues, the representative replica broadcasts a message $a$ consisting of a payment together with the dependencies accumulated by the issuer of the payment since the last broadcast. 
More precisely, when a replica receives a payment from a client, it executes the steps outlined in Listing \ref{lst:replica-brb}.

\begin{lstlisting}[
  caption = {\textbf{Using dependencies in \sysnamesig .} Representative replica broadcasts payment with dependencies.},
  label={lst:replica-brb},
  language=Astro,
  float,
  firstnumber=last,
]
@executes at the representative replica
@local state: DepMap deps[..] //dependencies per client

callback receive(a):
  let a be <Alice, n, b, x>
  Broadcast(<Alice, n, b, x, deps[Alice]>) 
  deps[Alice] := {}
\end{lstlisting}

In the life-cycle of a payment, the dependencies of the spender (Alice) are materialized into balance in her replicated \logname, while the payment itself becomes a new dependency for the beneficiary (Bob).
In particular, we obtain a full picture of the system by just re-defining the original approval and settling procedures of \Cref{lst:approve-tot} and \ref{lst:settle-tot} with \Cref{lst:approve-notot} and \ref{lst:settle-notot}, respectively.
Finally, \Cref{lst:deliver-unicast} shows how to handle the delivery of a proof at the representative replica of the beneficiary, which happens after a payment is settled.

\begin{lstlisting}[
  caption = {\textbf{Payment approval for \brbroadcast of \Cref{lst:brb-sig}.} Every replica executes this to approve a payment \emph{a}, assuming spender Alice.},
  label={lst:approve-notot},
  firstnumber=last,
  float,
  language=Astro,
 ]
func approve(a)
  let a be <Alice, n, _, x, dependencies>
  wait until sn[Alice] = n - 1
\end{lstlisting}

\begin{lstlisting}[
  caption = {\textbf{Payment settling procedure for \brbroadcast of \Cref{lst:brb-sig}.} Each replica executes this protocol to transition a payment \emph{a} to the final, settled state.},
  label={lst:settle-notot},
  firstnumber=last,
  float,
  language=Astro,
]
@executes at all system replicas
// Used dependencies per client
@local state: DepMap usedDeps[..] 

func settle(a)
  let a be <Alice, n, b, x, dependencies>
  
  // Keep only the never seen before dependencies
  newDeps = set(dependencies) \ usedDeps[Alice]
  usedDeps[Alice] = usedDeps[Alice] (*@$\cup$@*) newDeps
  
  bal[Alice] += balanceOf(newDeps) // Credit balance
  if bal[Alice] (*@<@*) x: return
    
  bal[Alice] -= x // Withdraw from Alice's balance
  sn[Alice] += 1
  xlogs[Alice].append(a)

  d = (Alice, n, b, x)
  // Send proof to Bob's representative (Credit message)
  trigger unicast(b, (d, Sign(d)))
\end{lstlisting}

\begin{lstlisting}[
  caption = {\textbf{Handling of dependencies for \brbroadcast of \Cref{lst:brb-sig}.} The representative replica executes this protocol every time a proof of an incoming payment is received.},
  label={lst:deliver-unicast},
  firstnumber=last,
  float,
  language=Astro,
]
@executes at the representative replica
@local state: DepMap deps[..] // dependencies per client
              DepMap partialDeps[..]

callback DeliverUnicast(proof)
  let proof be <payment, sig>
  let payment be <Alice, n, b, x>
  
  if !check(proof, payment): 
    return
  partialDeps[payment].add(proof)
  
  // An incoming payment that collects f+1 proofs becomes
  // a dependency.
  if len(partialDeps[payment]) = f + 1:
    deps[Alice].add(partialDeps[payment])
    delete(partialDeps[payment])
\end{lstlisting}


\end{document}